\documentclass[oneside]{article}
\usepackage[T1]{fontenc}
\usepackage{authblk}
\usepackage{float}
\usepackage{amsmath}
\usepackage{graphicx}
\usepackage{xfrac}
\usepackage[english]{babel}
\usepackage{lmodern}
\usepackage[T1]{fontenc}
\usepackage[latin9]{inputenc}
\PassOptionsToPackage{normalem}{ulem}
\usepackage{ulem}
\usepackage{mathtools}
\usepackage{graphicx}
\usepackage{esint}
\usepackage[unicode=true,
 bookmarks=false,
breaklinks=false,pdfborder={0 0 1},backref=false,colorlinks=false]
 {hyperref}
\usepackage[a4paper, total={210mm,297mm},margin=2.0cm]{geometry}
\usepackage{breakurl}
\usepackage[normalem]{ulem}

\usepackage{datetime}
\newdate{date}{16}{1}{2019}

\title{Disordered interfaces in soft fluids with suspended colloids}

\author[1]{Adriano Tiribocchi \thanks{Electronic address: \texttt{adriano.tiribocchi@iit.it}; Corresponding author}}
\author[2]{Marco Lauricella}
\author[2]{Andrea Montessori}
\author[3]{Simone Melchionna}
\author[1,2,4]{Sauro Succi}

\affil[1]{Center for Life Nano Science@La Sapienza, Istituto Italiano di Tecnologia, I-00161 Roma, Italy}
\affil[2]{Istituto per le Applicazioni del Calcolo CNR, Via dei Taurini 19, I-00185 Rome, Italy}
\affil[3]{Istituto dei Sistemi Complessi CNR, Consiglio Nazionale delle Ricerche, Dipartimento di Fisica, University of Rome La Sapienza, P.le A. Moro 2, I-00185 Rome, Italy}
\affil[4]{Harvard Institute for Applied Computational Science, Cambridge, Massachusetts 02138, USA}

\date{\displaydate{date}}

\begin{document}

\maketitle

\begin{abstract}
Computer simulations of bi-continuous two-phase fluids with intersparsed dumbbells show that, unlike rigid colloids, soft  dumbbells do not lead to arrested coarsening. However, they significantly alter the curvature dynamics of the fluid-fluid interface, whose probability density distributions are shown to exhibit (i) a universal spontaneous transition from an initial broad-shape distribution towards a highly localized one and (ii) super-diffusive dynamics with long-range effects. Both features may prove useful for the design of novel families of soft porous materials. 
\end{abstract}


\section{Introduction}

In the last few decades, outstanding progress in soft matter physics has opened up new routes for the design and manufacturing of innovative complex materials with bespoken properties\cite{Nieves,Piazza}. Of particular relevance are soft-glassy materials, such as foams and gels, which have found extensive use in many branches of modern industry, such as chemical and food processing. These systems hold a great interest, as they exhibit highly complex non-equilibrium effects, such as yield-stress behaviour and long-time relaxation, which dramatically affect their mechanical and rheological properties\cite{Berhier,Addad}. A pivotal role, underneath such non-trivial behavior, is played by complex interfaces which interact with the surrounding material and affect its stability and morphology. Two remarkable examples in point are offered by bijels\cite{Stratford,Herzig} and Pickering emulsions\cite{Pickering}. While the former material consists of a {\it disordered} fluid-fluid interface, where a dense layer of colloids is jammed (the resulting macroscopic structure is similar to an amorphous soft-solid), the latter is a collection of droplets stabilised by colloids sitting on an {\it ordered} fluid-fluid interface. The degree of order of the interface, as well as the interplay of its mesoscopic structure with the particle size, are crucial parameters which control mechanical properties and design of such systems. 

In this work, we present a computational study, performed by means of large-scale Lattice Boltzmann Particle-Dynamics (LBPD) simulations, of a soft material made up of a bi-continuous fluid with colloidal dumbbells, anchored at the fluid-fluid interface. Like in a bijel, the interface exhibits a high degree of disorder, although, unlike the bijel case, the flexible particles do not arrest the coarsening of the fluid domains. Our results show that the interface curvature captures crucial dynamical properties of the system, potentially useful for a targeted design of new soft mesoscale materials. For instance, the probability density functions (pdfs) of the curvature of the fluid-fluid interface exhibit a spontaneous transition from an initial broad distribution, reflecting high-curvature stretches, towards a localised, steady-state distribution associated with the interface flattening. 

In addition, our results suggest that the dynamics of pdf's of the curvature may be described in terms of a fractional Fokker-Planck equation, pointing to a super-diffusive dynamics of the fluid-fluid interface, possibly associated with long-range interactions. Within such picture, a local perturbation would propagate faster along the interface than in the bulk fluid, pointing to the former as to a sort of high-speed communication network within the fluid.

In the next section, we briefly outline the numerical model adopted to simulate the system. Afterwards, we report the results on the dynamics of the pdf's of the interface curvature and, finally, we present some conclusive remarks.

\section{Model and method}

Here, we briefly summarise the numerical model, directing the reader to to Refs.\cite{Melchionna,Melchionna2} for further details.

The equations of motion governing the physics of the system are the continuity and the Navier-Stokes equations in the incompressible limit
\begin{eqnarray}
\nabla\cdot {\bf u}&=&0,\label{cont_eq}\\
\frac{\partial{\bf u}}{\partial t}+({\bf u}\cdot\nabla){\bf u}&=&-\frac{1}{\rho}\nabla P + \nu\nabla^2{\bf u} +{\bf F},\label{Nav_stok}
\end{eqnarray}
where ${\bf u}$ is the fluid velocity, $\rho$ is the fluid density, $P$ is the isotropic pressure, $\nu$ is the kinematic viscosity and ${\bf F}$ is a body force representing the contribution due to the deviatoric stress.  

These equations are solved via a hybrid LB-Immersed boundary method\cite{Succi2,Succi3,Chen,Bernaschi}, already successfully adopted to study blood flows\cite{Melchionna,Melchionna2}. Within this approach, Eqs. (\ref{cont_eq})-(\ref{Nav_stok}) are numerically solved by means of a Shan-Chen LB method\cite{Shan1,Shan2}, in which fluid-fluid and fluid-particle interactions are included via the forcing term ${\bf F}$.  The former interaction term accounts for the presence of fluid-fluid interfaces separating the two components of the binary fluid, while the latter one models the hydrodynamic coupling between each colloid and the surrounding fluid. This term, in turn, includes two further contributions modeling, respectively, the translational coupling and the solvation force between colloids and fluid. 

Our colloids are represented as dumbbells composed of two spherical beads of radius $R$, located at equilibrium distance $d$, and interacting through an elastic potential with elastic constant $K$ (in Fig. \ref{fig1} we show a sketch of the model). These parameters define the aspect ratio of the dumbbell, $A=(d+2R)/2R$. The beads of the dumbbell are modeled via an Immersed buondary method, in which positions and velocities are indicated with ${\bf R}_i$ and ${\bf V}_i$ (and $i=1,...,N_b$, with $N_b$ number of beads), and a spherically symmetric diffused shape is achieved by means of an appropriate shape function $\tilde{\delta}_{\alpha}$ defined on an underlying Eulerian grid indicated by $\alpha=x,y,z$.

Note that both translational coupling and solvation force can be written in terms of the shape function and of its gradient. The former one can be explicitly written as ${\bf F}_i^T=-\sum_x\gamma_T\tilde{\delta}({\bf x}-{\bf R}_i)[{\bf V}_i-{\bf u}({\bf x})]$, where $\gamma_T$ is a translational coupling coefficient and the sum is performed over the particle volumetric extension described by the shape function $\tilde{\delta}$. The solvation force is designed to attract one component of the binary fluid and to simultaneously repel the other one for each bead. This ensures that dumbbells remain anchored at the fluid-fluid interface (see the zoom in Fig. \ref{fig1}) as each bead is, all the time, surrounded by a different fluid component. Its functional form is given by ${\bf F}_i^S=\sum_k\sum_{\alpha}\Lambda_k\rho_{\alpha}^k\nabla\tilde{\delta}_i({\bf x})$, where $k=1,2$ indicates the two phases of the binary fluid and $\Lambda_k$ is a constant determining the strength of the interaction between each bead with each component of the fluid, while its sign indicates whether the interaction is attractive (positive) or repulsive (negative). The action of both such forces must be counterbalanced by the opposite reaction on the fluid side, obtained by summing ${\bf F}_i^T$ and ${\bf F}_i^T$ over all the $N_b$ beads. Finally, in order to avoid colloidal interpenetration, a repulsive interaction, described by a Weeks-Chandler-Andersen potential\cite{Weeks}, is introduced.

\section{Results}

Next, we discuss the numerical results of a phase-separating binary fluid with colloids, focussing the study on the dynamics of the fluid-fluid interface curvature observed for different values of particle volume fraction (defined as $V_f=\frac{\frac{4}{3}\pi R^32N_b}{L^3}$, with $L$ lattice size) and aspect ratio $A$.

\subsection{Curvature dynamics of fluid-fluid interfaces with colloids}

\begin{figure*}
\centerline{\includegraphics[width=1.0\textwidth]{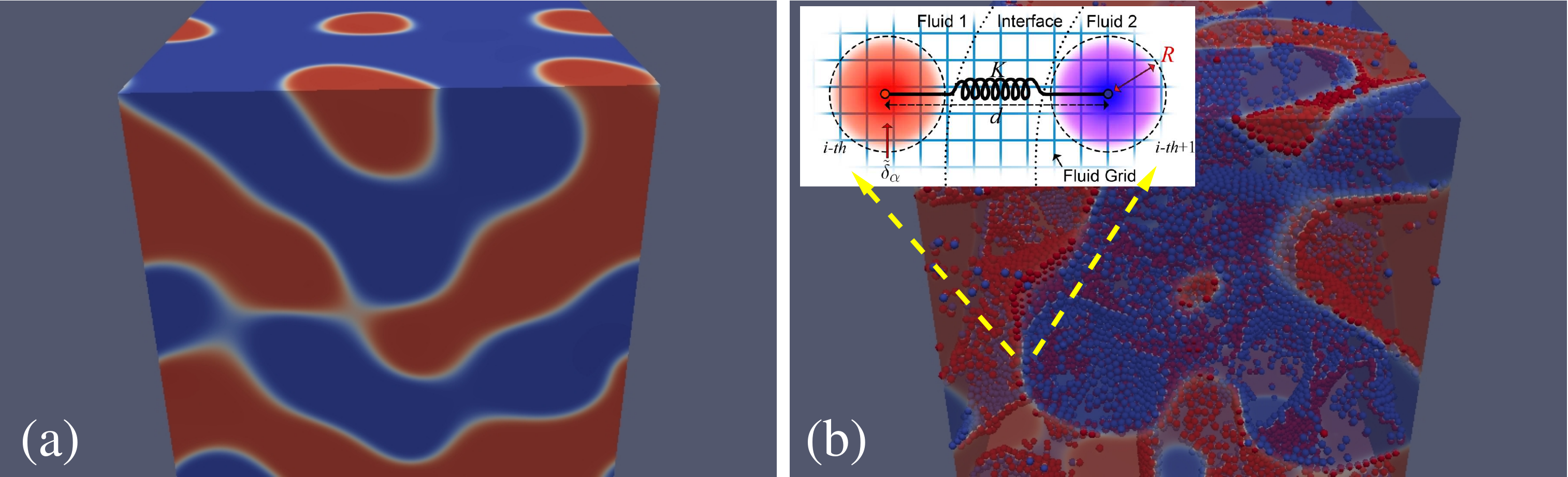}}
\caption{Two late time configuration (taken at $t=5500\Delta t$) of the density field when (a) $V_f=0$ and (b) $V_f\sim 0.036$, $A=2$. The zoom shows a cartoon of the colloidal dumbbell, anchored at the interface though a solvation force (see text). The beads (of predefined radius $R$) are modeled via an Immersed Boundary method, in which the spherical symmetry is achieved by using an {\it ad hoc} shape function $\tilde{\delta}_{\alpha}$ defined over an appropriate Eulerian mesh with Cartesian components $\alpha=x,y,z$. Beads are located at equilibrium distance $d$ and interact by means of an elastic potential, represented by a spring of elastic constant $K$. On the background the fluid lattice where lattice Boltzmann populations are defined.}
\label{fig1}
\end{figure*}

Simulations are performed on a cubic lattice of linear size $L=256$, with $\gamma_T=0.1$, and $\Lambda_k=\pm 0.1$. The binary fluid is initially in a disordered state, in which the two components are equally ($50:50$) mixed  (at a temperature $T$ over the critical value $T_c$ of the coexistence region of the phase diagram\cite{Bray}), and particles of predefined aspect ratio $A$ are randomly included in the lattice. Afterwards the mixture is quenched to a temperature $T<T_c$, in a region of the phase diagram where domains of the bicontinuous fluid gradually separate and grow with time. The equilibrium value of the fluid density $\rho^{eq}$ for the two fluid components (red and blu, see Fig. \ref{fig1}) can be found through simulations. We obtain $\rho^{eq}\simeq 0.1,1.9$, for the blue and the red one respectively; hence, the order parameter $\phi=(\rho_r-\rho_b)/(\rho_r+\rho_b)$ ranges roughly between $-0.9$ and $0.9$

In Fig. \ref{fig1}a-b we show two typical late-time configurations of the binary fluid density for $V_f=0$ and $V_f\sim 0.036$, $A=2$. Such value of $A$ sets the longitudinal length of the dumbbell particle at $\sim 8$ lattice points, comparable with the interface width, estimated around $5-7$ lattice spacings, as in previous works\cite{Melchionna,Melchionna2,Coveney}. 

In both cases, domains of each phase grow with time as the phase separation proceeds, until they achieve a sufficiently large size at late times, beyond which finite size effects become dominant. The inclusion of particles anchored at the fluid-fluid interface importantly alters domain morphology and interface curvature, as clearly visible by eye from our simulation results (see Fig. \ref{fig1}a and Fig. \ref{fig1}b). In particular, when $V_f=0$, isolated spherical domains emerge alongside large and amorphous ones with a flatter interface, while, when $V_f\neq 0$, large domains with higher interface curvature are observed. A natural question is then the following: what is the exact fluid-fluid interface dynamics? Understanding this, is crucial to correctly assess the different behaviour observed when colloids are included, and, more generally, the mechanical properties of the system.

Here we show that the fluid-fluid interface curvature represents a precious physical marker which can provide a quantitative response to that question. More specifically, the interface curvature, although rarely considered to evaluate the physics of a binary fluid \cite{Koga,Henry}, has been found capable to capture mechanical properties like inhomogeneities in the particle volume fraction and an (arguably) fractional interface dynamics, not easily inferable from experiments but of great importance for the design of soft porous materials.

\begin{figure*}
\centerline{\includegraphics[width=1.0\textwidth]{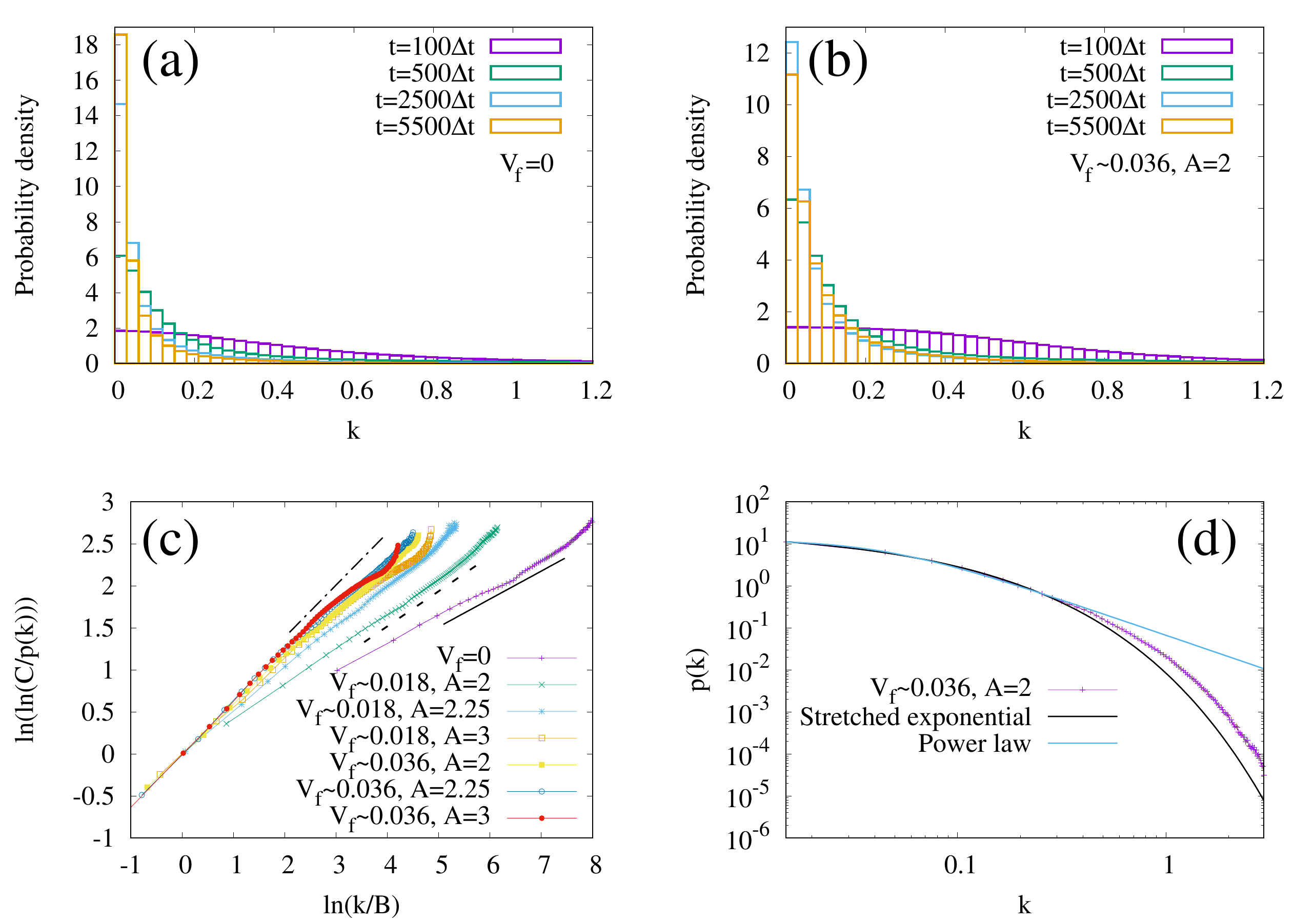}}
\caption{Time evolution of the probability density function of the curvature $k$ for (a) $V_f=0$ and (b) $V_f\sim 0.036$, $A=2$. In (c)  we show a fit of the steady state profiles of $p(k)$ (taken at $t=5500\Delta t$, the same simulation time of the configurations shown in Fig. \ref{fig1}) with a stretched exponential function $f(k)=Ce^{{-k/B}^{\theta}}$, where $C$, $B$ and $\theta$ are fitting parameters, for different values of particle volume fraction ($V_f=0$, $V_f\sim 0.018$ and $V_f\sim 0.036$) and particle aspect ratio ($A=2$, $A=2.25$ and $A=3$). For $V_f=0$ $\theta\simeq 0.33$. For $V_f\sim 0.018$ $\theta\simeq 0.41$ if $A=2$, $\theta\simeq 0.51$ if $A=2.25$ and $\theta\simeq 0.56$ if $A=3$. For $V_f\sim 0.036$ $\theta\simeq 0.58$ if $A=2$, $\theta\simeq 0.63$ if $A=2.25$ and $\theta\simeq 0.63$ if $A=3$.  Solid, dashed and dot-dashed lines indicate the slope of the purple (plusses) line, where $\theta\sim 0.33$, of the green (crosses) line, where $\theta\sim 0.41$, and of the red (filled circles) line, where $\theta\sim 0.63$. In (d) we compare a stretched exponential (black) with a power law (cyan) $g(k)\sim k^{\gamma}$ for $V_f\sim 0.036$, $A=2$, on a log-log scale. We got $\gamma\simeq -1.7$}
\label{fig2}
\end{figure*}

In Fig. \ref{fig2}a-b we show the time evolution of $p(k,t)$, the probability density function of the fluid-fluid interface curvature, for $V_f=0$ and $V_f\sim 0.036$, $A=2$. The curvature is defined as $k=|\nabla\cdot (\frac{\nabla\phi}{||\nabla\phi||})|$, in which there is no distinction between positive and negative sign. At early times ($t=100\Delta t$) $p(k,t)$ looks almost uniformly distributed as long as $k \leq 0.4$ (the fluid contains high-curvature stretches), and then gently decays to 0 for higher values of $k$. Afterwards, for $t \geq 500\Delta t$, the pdf collapses towards low values of $k$, as the fluid-fluid interface gets flatter, and quickly decays to $0$ for high values of $k$. The system exhibits  a distinct spontaneous transition from an initial mixed/disordered state, with broad-shaped pdfs, to a late-time ordered state with rather high localised distributions, a scenario broadly resembling that of the ``winner-takes-it-all'' complex systems\cite{Thurner}. Remarkably, this collapse-like dynamics occurs regardless of the particle volume fraction $V_f$ and particle aspect ratio $A$, except for the peak of $p(k)$, higher for $V_f=0$ as the interface becomes smoother (see, for instance, Fig.\ref{fig2}b where the time evolution is shown for $V_f\sim 0.036$, $A=2$). While, on the one hand, such transition exhibits a significant degree of universality, on the other hand, it is only weakly affected by the presence of colloids. However, their effect is better highlighted by looking at the steady state behavior of $p(k,t)$, whose study also provides insights into the structure of a statistical dynamic model of the interface dynamics (or equivalently into the nature of a kinetic equation for the pdf of the curvature). To this scope, it is useful to evaluate the functional form of $p(k,t)$ at the steady state. 



In Fig. \ref{fig2}c, we show the late time (and steady state, achieved for $t>5000\Delta t$) configurations of $p(k)$ for different values of $V_f$ and $A$, in the curvature space. We find that, as long as $k\leq 1$,  the relaxation dynamics can be well described in terms of a stretched exponential function in curvature space $p(k)=Ce^{-({k/B})^{\theta}}$, where $C$, $B$ and $\theta$ are fitting parameters. Our simulations show that $\theta$ is always substantially lower than $1$, with $\theta$ being $\sim 0.34$ for a binary fluid with $V_f=0$, and ranging from $\sim 0.45$ to $\sim 0.65$ when $V_f\neq 0$ (all values of $\theta$ are given in the caption of Fig.\ref{fig2}).

Given that  power laws and a stretched exponentials exhibit a similar behavior for low values of $k$ ($k<1$), one might wonder whether the relaxation dynamics of the interface curvature could also be fitted by a power law $k^{\gamma}$, In Fig.\ref{fig2}d, we present a fit of $p(k)$ for $V_f\sim 0.036$, $A=2$ with both functions. From this figure, it is seen that they both provide a reasonable approximation of $p(k)$ for $k<1$, whereas the decaying behavior at high values of $k$ is better captured (although not perfectly) by the stretched exponential.

This behavior supports the idea that the fluid-fluid interface acts as a sort of dynamic network, supporting long-range interactions within the fluid system. This means that a local perturbation would propagate more rapidly along the interface rather than in the bulk fluid, pointing to the interface as to a sort of self-consistent privileged communication network, characterised by a higher longitudinal stiffness than the bulk. At the moment, this is however only a speculation, which we plan to test for the future, by setting up specific probes for the propagation speed of density perturbations along the interface.

From a mathematical standpoint, this dynamics points to a fractional Fokker-Planck for $p(k,t)$, of the form\cite{Risken}
\begin{equation}\label{fract}
 \frac{\partial}{\partial t}p(k,t)={_0}D_t^{1-\theta}\left(\frac{\partial}{\partial k}V^{\prime}(k)+D_{\theta}\frac{\partial^2}{\partial k^2}\right)p(k,t),
\end{equation}
where $V(k)$ is a potential associated to a drift force, $D_{\theta}$ is a generalized diffusion constant and ${_0}D_t^{1-\theta}\equiv (d/dt(_0D_t^{-\theta}))$ is the fractional Riemann-Liouville operator, basically a memory kernel decaying like $1/(t-t')^{1-\theta}$ \cite{Riemann}. 

Based on the theory of fractional Fokker-Planck equation, it is known that a power law behaviour of $p(k)$ would associate with an underlying Levy flight\cite{Zaslavsky}, in which, pictorially, the propagation of a local perturbation would occur through an alternate sequence of trapping intervals, followed by anomalously long ``flights'', leading  to super-diffusive dynamics. Even though this is not exactly our case (see Fig\ref{fig2}c), a stretched exponential suggests nonetheless the existence of a super-diffusive curvature dynamics with long-range effects.

\section{Conclusions}

To summarize, we have presented a computational study, performed by means of large-scale Lattice Boltzmann Particle-Dynamics simulations, on the dynamics of the fluid-fluid interface in a bicontinuous fluid in the presence of elastic colloidal particles. These particles (dumbbells), modeled via an immersed boundary method, are anchored at the fluid-fluid interface due to a solvation force. 

We show that the curvature of the interface is a key observable, capturing relevant dynamic properties of the system which are not revealed by standard coarsening indicators. In particular, it is found that  that the probability density functions of the curvature exhibit a spontaneous transition from an early broad distribution to a late-time localized unimodal one. In addition, at steady state, the pdfs follow a stretched exponential behaviour, whose exponent is significantly lower than $1$ and sensitive to changes in the particle aspect ratio and particle volume fraction. This behaviour suggests that the fluid-fluid interface may support long-range correlations, whose dynamics may be described by a fractional Fokker-Planck equation or an equivalent fractional Brownian motion.

Besides a theoretical interest on its own, this study may also offer practical hints for the experimental design of novel mesoscale porous materials with innovative mechanical properties relevant, for example, to tissue engineering and biomedical applications. Future experimental tests might open new avenues and stimulate further questions relevant to the technology of new families of soft porous materials.

\section*{Acknowledgments}
The authors acknowledge funding from the European Research Council under the European Union's Horizon 2020 Framework Programme (No. FP/2014-2020) ERC Grant Agreement No.739964 (COPMAT).

\end{document}